\documentclass[12pt]{article}
\usepackage{latexsym}

\begin{document}
\vspace*{-.6in}
\thispagestyle{empty}
\begin{flushright}
CALT-68-2472
\end{flushright}
\baselineskip = 18pt

\vspace{1.5in} {\Large
\begin{center}
THE THREE-STRING VERTEX \\ FOR A PLANE-WAVE
BACKGROUND\end{center}} \vspace{.5in}

\begin{center}
John H. Schwarz
\\
\emph{California Institute of Technology\\ Pasadena, CA  91125, USA}
\end{center}
\vspace{1in}

\begin{center}
\textbf{Abstract}
\end{center}
\begin{quotation}
\noindent The three string vertex for Type IIB superstrings in a
maximally supersymmetric plane-wave background can be constructed
in a light-cone gauge string field theory formalism. The detailed
formula contains certain Neumann coefficients, which are functions
of a momentum fraction $y$ and a mass parameter $\mu$. This paper
reviews the derivation of useful explicit expressions for these
Neumann coefficients generalizing flat-space ($\mu =0$) results
obtained long ago. These expressions are then used to explore the
large $\mu$ asymptotic behavior, which is required for comparison
with dual perturbative gauge theory results. The asymptotic
formulas, exact up to exponentially small corrections, turn out to
be surprisingly simple.
\end{quotation}

\vspace{1.0 in}

\centerline{\it Dedicated to the Memory of Ian Kogan}

\newpage

\pagenumbering{arabic}

\section{Introduction}

A maximally supersymmetric plane-wave background in ten dimensions
is an exact solution of Type IIB superstring theory
\cite{Blau:2001ne}. Moreover, the string theory in this background
is tractable, despite the fact that the background contains a
nonvanishing RR field, provided that one uses the light-cone gauge
Green--Schwarz (GS) formalism \cite{Metsaev:2001bj}. In that
approach the world-sheet theory consists of free massive bosons
and fermions. Thus it is trivial to read off the complete spectrum
of the noninteracting theory.

The string states and their interactions are holographically dual
to certain operators and their correlation functions in ${\cal N}
= 4$ super Yang-Mills theory in a suitable limit
\cite{Berenstein:2002jq}. However, this paper will only consider
the string side of the story and not discuss the duality.

The string interactions are encoded (using the formalism of
light-cone-gauge string field theory) in a cubic interaction
vertex. The three-string vertex has been formulated by Spradlin
and Volovich \cite{Spradlin:2002ar}\cite{Spradlin:2002rv} and
explored further by other authors \cite{Pankiewicz:2002tg}. These
results generalize the flat-space light-cone gauge field theory
results of \cite{Green:1982tc}\cite{Green:hw} to the plane-wave
geometry. For recent reviews see
\cite{Pankiewicz:2003pg}--\cite{Sadri:2003pr}.

This paper reviews work that makes the formulas for the Neumann
coefficients that enter in the interaction vertex more explicit.
These coefficients are defined in the first instance in terms of
the inverse of a certain infinite dimensional matrix. The first
step is to express this inverse matrix in terms of a certain
infinite component vector
\cite{Schwarz:2002bc}\cite{Pankiewicz:2002gs}. The next step is to
derive an expression for this vector in terms of a certain scalar
quantity and then to derive an explicit formula for the scalar
\cite{He:2002zu}. Having obtained useful expressions for the
Neumann coefficients (and hence the three superstring vertex), one
can then explore the large $\mu$ (large curvature) limit, which is
required for making contact with dual perturbative gauge theory
computations.

\section{Review of basic formulas}

The type IIB superstring in the maximally supersymmetric
plane-wave background is described in light-cone gauge by a free
world-sheet theory. The eight bosonic and eight fermionic
world-sheet fields each have mass $\mu$, a parameter that enters
in the description of the plane-wave geometry and the RR five-form
field strength. The mass term has two important consequences. One
is that it leads to a mixing of left-movers and right-movers. The
other is that the zero modes are also described by harmonic
oscillators of finite frequency. Altogether, a convenient
labelling of the bosonic lowering and raising operators arising
from quantization of the free world-sheet theory is $a_m^I$ and
$a_m^{I \dagger}$, where $m$ runs from minus infinity to plus
infinity and $I = 1,\ldots,8$. These satisfy ordinary oscillator
commutation relations
\begin{equation}
[a_m^I, a_n^{J \dagger}] = \delta_{mn} \delta^{IJ}.
\end{equation}
There are also fermionic oscillators $b_m^{\alpha}$ and
$b_m^{\alpha \dagger}$, which will not be discussed in this paper.

The spectrum of the free string theory is described by the
light-cone Hamiltonian
\begin{equation}
H_2 = \sum_{m= -\infty}^{\infty} \omega_m N_m ,
\end{equation}
where $N_m$ is the number of excitations of level $m$ oscillators
\begin{equation}
N_m = \sum_{I=1}^8 a_m^{I \dagger} a_m^I + {\rm fermionic\, terms}
,
\end{equation}
and the frequencies are given by
\begin{equation}
\omega_m = \sqrt{m^2 + \mu^2 \alpha^2}.
\end{equation}
The second term in the square root is actually $(\alpha' \mu
p_-)^2$, but we  define $\alpha = \alpha' p_-$. (In flat space, we
used the symbol $p^+$ rather than $p_-$, but in curved space a
subscript is more natural, since momenta are conjugate to
coordinates that are defined with superscripts.) The physical
spectrum is given by the product of all the oscillator spaces
subject to one constraint
\begin{equation}
\sum_{m= -\infty}^{\infty} m N_m =0.
\end{equation}
In flat space this constraint reduces to the usual level-matching
condition for left-movers and right-movers.

The three-string interaction vertex for type IIB superstrings in
flat space was worked out in \cite{Green:1982tc}\cite{Green:hw}
and generalized to the plane-wave geometry in
\cite{Spradlin:2002ar}\cite{Spradlin:2002rv}\cite{Pankiewicz:2002tg}.
The formula can be written rather elegantly in terms of
functionals, but to make its meaning precise and easily applicable
to specific external states, it is desirable to expand it out in
terms of oscillators. A convenient notation uses a tensor product
of three string Fock spaces, labeled by an index $r=1,2,3$. Then
the three string interaction vertex contains a factor
\begin{equation}
|\, V_B\rangle = {\rm exp} \left( {1\over 2} \sum_{r,s =1}^3 \sum_{m,n
=-\infty}^{\infty} \sum_{I=1}^8 a_{mr}^{I\dagger} \bar N_{mn}^{rs}
a_{ns}^{I\dagger}\right) |\, 0 \, \rangle.
\end{equation}
The quantities $\bar N_{mn}^{rs}$, called Neumann coefficients,
are the main objects of concern in this paper.\footnote{Their definition
here differs from that used in \cite{Green:1982tc}\cite{Green:hw}
by factors of $\sqrt{m n}$. The definition given
here is more natural for the $\mu\neq 0$ generalization.} The three
string vertex also contains a similar expression $|\, V_F\rangle$ made
out of the fermionic oscillators and a ``prefactor'' that is
polynomial in the various oscillators. We will not consider either
of these in this paper. Suffice it to say that they are made out of the same
basic objects, so that the results described here for the the bosonic
factor are applicable to them as well.

In describing the Neumann matrices, it is convenient to consider
separately the cases in which each of the indices $m,n$ are either
positive, negative or zero. Henceforth, the symbols $m,n$ will
always denote positive integers. Negative integers will be
indicated by displaying an explicit minus sign.  One result of
\cite{Spradlin:2002ar}, for example, using matrix notation for the
blocks with positive indices, is
\begin{equation}
\bar N^{rs} = 1 - 2 (C_r C^{-1})^{1/2} A^{(r)T}
\Gamma_+^{-1}A^{(s)} (C_s C^{-1})^{1/2}.
\end{equation}
Here $C_{mn} = m \delta_{mn}$ and $(C_r)_{mn} = \omega_{rm}
\delta_{mn}$, where
\begin{equation}
\omega_{rm} = \sqrt{ m^2 +
(\mu\alpha_r)^2}.
\end{equation}
The definitions of $A^{(r)}$ and $\Gamma_+$, and other expressions
that appear here, are collected in Appendix A.

The blocks with both indices negative are related in a simple way
to the ones with both indices positive by
\begin{equation}
\bar N_{-m-n}^{rs} = - \left( U_r \bar N^{rs} U_s \right)_{mn},
\end{equation}
where
\begin{equation}
U_r = C^{-1} (C_r - \mu \alpha_r).
\end{equation}
In the case of $\bar N^{33}$ these are the only nonvanishing
terms. For the other Neumann coefficients the other nonvanishing
terms are
\begin{equation} \label{Nmzero}
\bar{N}_{m0}^{rs} = \bar{N}_{0m}^{sr} = \sqrt{2\mu\alpha_s}
\epsilon^{st}\alpha_t [(C_r C^{-1})^{1/2} A^{(r)T} Y]_m \quad r= 1,2,3 \quad s = 1,2
\end{equation}
\begin{equation}
\bar{N}_{00}^{rs} = (1 + \mu \alpha k) \epsilon^{rt}
 \epsilon^{su} \sqrt{\alpha_t \alpha_u} \quad r,s = 1,2
\end{equation}
\begin{equation}
\bar{N}_{00}^{r3} = \bar{N}_{00}^{3r} = - \sqrt{\alpha_r}
\quad r = 1,2.
\end{equation}
Here we have introduced (see Appendix A)
\begin{equation}
Y = \Gamma_+^{-1} B,
\end{equation}
\begin{equation}
k = B^T \Gamma_+^{-1} B,
\end{equation}
\begin{equation}
y = - \alpha_1/\alpha_3,
\end{equation}
and (setting $\alpha_3 = -1$)
\begin{equation}
\alpha = \alpha_1\alpha_2\alpha_3 = -y(1-y).
\end{equation}
The asymmetry between string number three and the other two
strings is a reflection of the fact that the $\mu$ dependence of
the formula breaks the cyclic symmetry that is present in the flat
space case.

To make the formulas useful for comparison with the dual gauge
theory, it would be helpful to have explicit formulas for the
various matrix multiplications and inversions that appear. The
quantities that we especially would like to evaluate explicitly
are the matrix $\Gamma_+^{-1}$, the vector $Y =\Gamma_+^{-1} B$,
and the scalar $k = B^T \Gamma_+^{-1} B$. In the case of flat
space ($\mu =0$) the results are known. Specifically
\begin{equation} \label{factorize}
\bar N_{mn}^{rs} = - {mn\alpha \over m\alpha_s + n \alpha_r} \bar
N_m^r \bar N_n^s \quad {\rm for}\,  \mu =0
\end{equation}
where
\begin{equation} \label{Nrflat}
\bar N_m^r = {\sqrt{m} \over \alpha_r} f_m(-\alpha_{r+1}/\alpha_r)
e^{m\tau_0 /\alpha_r} \quad {\rm for}\,  \mu =0,
\end{equation}
\begin{equation}
f_m(\gamma) = {\Gamma(m\gamma) \over m!\, \Gamma (m\gamma +1 -m)},
\end{equation}
and
\begin{equation}
\tau_0 = \sum_{r=1}^3 \alpha_r \, {\rm log} |\alpha_r| .
\end{equation}
In particular, still for $\mu =0$, $\Gamma_+^{-1} = {1\over 2} (1
- \bar N^{33})$, $Y_m = - \bar N_m^3$, and $k = 2\tau_0/\alpha$.

\section{Factorization theorem}

In this section we will derive the generalization of
eq.~(\ref{factorize}) that holds for the plane-wave geometry. The
method of derivation is a fairly straightforward generalization of
the one used for flat space in \cite{Green:1982tc}. We begin by
defining
\begin{equation} \label{tildeG}
\tilde\Gamma_+ = \sum_{r=1}^3 A^{(r)} U_r^{-1} A^{(r)T},
\end{equation}
which differs from the $\Gamma_+$ by the replacement of $U_r$ by
$U_r^{-1}$. Then we consider the product
\begin{equation}
\Gamma_+\,  C^{-1}\,  \tilde\Gamma_+ = (U_3 + \sum_1^2 A^{(r)} U_r
A^{(r)T})\, C^{-1}\, (U_3^{-1} + \sum_1^2 A^{(s)} U_s^{-1}
A^{(s)T})
\end{equation}
Using various identities given in Appendix A, this simplifies to
\begin{equation}
\Gamma_+\,  C^{-1}\,  \tilde\Gamma_+ = U_3 C^{-1} \tilde\Gamma_+ +
\Gamma_+ C^{-1} U_3^{-1} -{1\over 2} \alpha_1 \alpha_2 B B^T.
\end{equation}
The next step is to use eqs.~(\ref{Uformula}) and (\ref{AAT2}) to
deduce that
\begin{equation}
\tilde\Gamma_+ = \Gamma_+ + \mu \alpha BB^T
\end{equation}
Substituting this into the previous equation and multiplying left
and right by $\Gamma_+^{-1}$ gives
\begin{equation} \label{idwithZ}
C^{-1} U_3^{-1} \Gamma_+^{-1} + \Gamma_+^{-1} U_3 C^{-1} = C^{-1}
+ {1\over2} \alpha_1 \alpha_2 Y Y^T + \mu \alpha Z Y^T
\end{equation}
where we have defined
\begin{equation}
Z = (1 -\Gamma_+^{-1} U_3) C^{-1} B.
\end{equation}

The next step is to eliminate $Z$ from eq.~(\ref{idwithZ}). This
is achieved by multiplying the equation on the right with the
vector $B$. This gives a linear equation for $Z$, whose solution
is
\begin{equation}
Z = {1\over 1 + \mu \alpha k}(C^{-1} U_3^{-1} - {1\over 2}
\alpha_1 \alpha_2 k)  Y.
\end{equation}
Substituting this back into eq.~(\ref{idwithZ}) and simplifying
gives the formula
\begin{equation} \label{symformula}
\{ \Gamma_+^{-1} , C_3\} = C + {1\over2} {\alpha_1 \alpha_2 \over
1 + \mu \alpha k} C U_3^{-1} Y Y^T CU_3^{-1}.
\end{equation}
In terms of components
\begin{equation} \label{Ginv}
(\Gamma_+^{-1})_{mn} = \frac{m}{2\omega_m} \delta_{mn}
+\frac{y(1-y) (\omega_m - \mu)(\omega_n - \mu)
Y_m Y_n}{2[1 - \mu y (1-y)k]
(\omega_m + \omega_n)},
\end{equation}
where
\begin{equation}
\omega_m = \omega_{3m} = \sqrt{m^2 + \mu^2} .
\end{equation}

This result be recast as a formula for the Neumann coefficient matrix $\bar
N^{33}_{mn}$. The result is
\begin{equation}
\bar N^{33}_{mn} = - {mn\alpha_1 \alpha_2 \over 1 + \mu \alpha k}
{\bar N_m^3 \bar N_n^3 \over \omega_{3m} + \omega_{3n}}
\end{equation}
where
\begin{equation}
\bar N_m^3 = -\left[ (C^{-1}C_3)^{1/2} U_3^{-1} Y\right]_m.
\end{equation}
Some further simple manipulations give the generalization
\cite{Schwarz:2002bc}\cite{Pankiewicz:2002gs}
\begin{equation} \label{Nfactor}
\bar N^{rs}_{mn} = - {mn\alpha \over 1 + \mu \alpha k} {\bar N_m^r
\bar N_n^s \over \alpha_s\omega_{rm} + \alpha_r\omega_{sn}}
\end{equation}
where
\begin{equation} \label{Nmr}
\bar N_m^r = -\left[ (C^{-1}C_r)^{1/2} U_r^{-1} A^{(r)T}
Y\right]_m.
\end{equation}
This is the desired generalization of the flat-space formula
eq.~(\ref{factorize}). However, we still require a
generalization of eq.~(\ref{Nrflat}) as well as
an explicit formula for $k$. Note that combining eq.~(\ref{Nmr})
with eq.~(\ref{Nmzero}) gives
\begin{equation} \label{Nmzerors}
\bar{N}_{m0}^{rs} = \bar{N}_{0m}^{sr} = - \sqrt{2\mu\alpha_s}
\epsilon^{st}\alpha_t U_r \bar N_m^r \quad r= 1,2,3 \quad s = 1,2.
\end{equation}

\section{Determination of $Y$ and $k$}

To complete the explicit determination of the Neumann matrices,
and thus the three string vertex, we need useful formulas for
\[(\Gamma_+^{-1})_{mn},\quad  Y_m =
(\Gamma_+^{-1} B)_m,\quad k = B^T \Gamma_+^{-1} B\]
as functions of $y$ and $\mu$. In view of of eq.~(\ref{Ginv}),
if we knew $Y_m$ and $k$ we would know
$(\Gamma_+^{-1})_{mn}$. The strategy that we
will use for obtaining them is to derive first-order
differential equations (in $\mu$) and input the known values at
$\mu = 0$ as initial conditions.
Using the various definitions and identities in
Appendix A, ref.~\cite{He:2002zu} derived the differential equation
\begin{equation} \label{diffeq}
\frac{\partial Y_m}{\partial \mu} = \left[\frac{1}{2} \frac{\partial
F}{\partial \mu} \left(1 - \frac{\mu}{\omega_m}\right) -
\frac{\mu}{\omega_m^2}\right] Y_m ,
\end{equation}
where
\begin{equation} \label{Fdef}
F (\mu, y) = \log [1 - \mu y (1-y) k (\mu, y)].
\end{equation}
The derivation of eq.~(\ref{diffeq}) is sketched in Appendix B.

Eq.~(\ref{diffeq}) has the solution
\begin{equation} \label{Ysoln}
Y_m (\mu, y) = \frac{m}{\omega_m} \exp \left[\frac{1}{2} \int_0^{\mu}
\frac{\partial F}{\partial \mu} \left(1 - \frac{\mu}
{\omega_m}\right)d\mu\right] Y_m (0, y) .
\end{equation}
Thus, since $Y_m (0, y)$ is known, if we knew $F(\mu, y)$, we would know
$k(\mu, y)$ and $Y_m (\mu,y)$ and hence all the Neumann coefficients.
Since, we have one fewer equations than unknowns, we need to input one additional
piece of information. One that is easy to obtain and does the job is
the asymptotic formula
\begin{equation}
Y_m \sim \frac{m}{2\mu} B_m + O(\mu^{-2})
\end{equation}
for large $\mu$. Combining this condition with eq.~(\ref{Ysoln}) implies that
\begin{equation}
\exp \left\{ \frac{1}{2} \int_0^\infty \frac{\partial F}{\partial \mu}
\left(
1 - \frac{\mu}{\omega_m}\right) d\mu \right\} =
\frac{B_m}{2Y_m(0)}.
\end{equation}
Taking the logarithm,
integrating by parts, and substituting the known value of the right-hand side gives
\begin{equation} \label{inteqn}
 \int_0^\infty (m^2 + \mu^2)^{-3/2} F(\mu, y) d\mu =
G(m,y),
\end{equation}
where
\begin{equation} \label{Gchoice}
G(z,y) = {2\tau_0 \over z} + {2 \over z^2} {\rm log} \left(
{\Gamma(1+z)\over \Gamma( 1+ zy) \Gamma(1+ z(1-y))} \right).
\end{equation}

This formula must hold for $m = 1,2, ...$ and $0 \leq y \leq 1$.
The inverse integral transform, which does not seem to exist in
the mathematical literature, determines $F(\mu, y)$. Ref.~\cite{He:2002zu}
proves that
for a function $G(m,y)$ that is holomorphic
in the right half $m$ plane
and vanishes at infinity in that half plane, the inverse integral transform that solves
the integral equation eq.~(\ref{inteqn}) is
\begin{equation}
F(\mu ,y) = - {i\mu^2 \over \pi} \int_0^\pi {\rm cos}\, \theta ~
G(-i\mu {\rm cos}\, \theta, y) d\theta .
\end{equation}
The proof is sketched in Appendix~C.
Using this, one can show that for our specific choice of
$G(m, y)$ in eq.~(\ref{Gchoice})
\begin{equation}
F(\mu ,y) = 2\mu \tau_0 + 2\sum_{r=1}^3 \phi(\mu \alpha_r),
\end{equation}
where
\begin{equation}  \label{phieqn}
\phi (x) = \sum_{n=1}^{\infty} \left[ {\rm log}\left(  { \sqrt{n^2 + x^2}
+x \over n} \right) - { x \over n} \right].
\end{equation}

We now have sufficiently explicit formulas to construct large
$\mu$ asymptotic expansions.\footnote{This analysis was initiated
in \cite{Klebanov:2002mp} and completed in \cite{He:2002zu}.}
Large $\mu$ corresponds to small $\lambda'=1/\mu^2$, which is the
effective coupling constant in the dual gauge theory
\cite{Berenstein:2002jq}. Differentiating eq.~(\ref{phieqn})
twice, introducing a contour integral (Sommerfeld-Watson)
representation of the series, and integrating by parts one can
show that
\begin{equation}
\phi''(x) = - x \sum_{n=1}^\infty \frac{1}{(x^2 + n^2)^{3/2}}
= - \frac{1}{x} + \frac{1}{2 x^2} - \pi \int_1^\infty
\frac{z dz}{\sqrt{z^2-1}} \frac{1}{\sinh^2(\pi x z)}.
\end{equation}
Integrating back, this allows us to deduce that
\begin{equation}
F(\mu,y) = - \ln [4 \pi \mu y(1-y)]
+ J(\mu y) + J(\mu(1-y)) - J(\mu),
\end{equation}
where
\begin{equation}
J(x) = \frac{2}{\pi} \int_1^\infty \frac{\ln(1 - e^{-2 \pi x z})
}{z \sqrt{z^2 - 1}} dz.
\end{equation}
Note that $J \sim \exp(-2\pi x)$ for large $x$.

\section{Asymptotic behavior}

We can now derive asymptotic results
that include all inverse powers of $\mu$ and have leading
corrections of order ${\rm exp} [-2\pi y \mu]$
(if $y\leq 1/2$). For example,
\begin{equation} \label{Fasy}
F(\mu,y) \approx - \ln [ 4 \pi \mu y (1-y)].
\end{equation}
The amazing thing is that a remarkably simply expression captures
the result to all finite orders in $1/\mu$ and therefore one can
easily read off predictions for the dual gauge theory that are
valid (at this order of nonplanarity) to all orders in
perturbation theory! The exponentially suppressed corrections
encoded by the $J$ functions correspond to nonperturbative effects
in the dual gauge theory. What these effects are, however, is
mysterious. They don't seem to be related to instantons or any
other familiar nonperturbative phenomena.

Combining eqs.~(\ref{Fasy}) and(\ref{Fdef}) one finds that
\begin{equation}
\label{kexpansion}
k(\mu,y) \approx \frac{1}{\mu y(1-y)} - \frac{1}{4 \pi \mu^2 y^2 (1-y)^2}.
\end{equation}
Similarly, substituting eq.~(\ref{Fasy}) in eq.~(\ref{Ysoln})
gives
\begin{equation}
\label{finaly} Y_m(\mu,y) \approx \sqrt{\frac{\mu+\omega_m}{2\mu}}
\frac{m}{2 \omega_m} B_m ,
\end{equation}
or in matrix notation
\begin{equation}
\label{yapprox}
Y \approx \frac{1}{2 \sqrt{2 \mu}} U_3^{1/2} C^{3/2} C_3^{-1}  B.
\end{equation}
From this it follows that
\begin{equation}
\label{finalnv} \bar N^r_n \approx \frac{(-1)^{r(n+1)} }{2 \pi
y(1-y)} \sqrt{  \frac{|\alpha_r|}{2 \mu n \omega_{rn} U_{rn}} }
s_{rn}, \qquad r \in \{1,2,3 \}
\end{equation}
where
\begin{equation}
s_{1m} = s_{2m} = 1, \qquad s_{3m} = - 2 \sin(\pi m y).
\end{equation}
The asymptotic expansions of the Neumann matrices are then given by
substitution in eqs.~(\ref{Nfactor}) and (\ref{Nmzerors}).

\section*{Acknowledgments}
I am grateful to M. Spradlin and A. Volovich for helpful discussions.
This work was supported in part by the U.S. Dept. of
Energy under Grant No. DE-FG03-92-ER40701.

\section*{Appendix A. Useful definitions and identities}

The light-cone momenta in the minus direction that appear
in the three-string vertex are defined to
be $\alpha_r/\alpha'$, $r=1,2,3$. For the process in which string \#3 splits into
strings \#1 and \#2, we take $\alpha_1, \alpha_2 > 0$, $\alpha_3 <
0$. Momentum conservation implies that $\alpha_1 + \alpha_2 + \alpha_3 =0$. 
We also define the momentum 
fraction carried by string \#1 to be $y
= -\alpha_1 / \alpha_3$, which satisfies $ 0 < y < 1$. It follows
that $1-y = -\alpha_2 / \alpha_3$ is the momentum fraction carried by
string \#2. It is sometimes
convenient to set $\alpha_3 = -1$, which can always be achieved by a 
suitable Lorentz boost. Only then does the mass parameter $\mu$ have an
invariant meaning.

The matrices $A^{(r)}_{mn}$, which appear in the Neumann
coefficients, are given by
\begin{equation}
A^{(1)}_{mn} = {2 \over \pi} (-1)^{m+n+1}  \sqrt{m n} {y \,
{\rm sin} (m \pi y) \over n^2 - m^2 y^2},
\end{equation}
\begin{equation}
A^{(2)}_{mn} = {2 \over \pi} (-1)^{m}  \sqrt{m n} {(1-y)\,
{\rm sin} (m \pi y) \over n^2 - m^2 (1-y)^2},
\end{equation}
and $A^{(3)}_{mn} = \delta_{mn}$. The indices $m,n$ range from 1
to infinity. Additional quantities that we need are
\begin{equation}
B_m = {2 \alpha_3 \over \pi \alpha_1 \alpha_2} (-1)^{m} { {\rm
sin} (m \pi y) \over m^{3/2} }
\end{equation}
and
\begin{equation}
C_{mn} = m \delta_{mn}.
\end{equation}
These quantities are all $\mu$ independent and were defined
already in the flat space analysis.

The infinite matrices $A^{(r)}_{mn}$ and the infinite vector $B_m$
satisfy a number of useful relations, which we record here
\begin{equation}
A^{(r) T} C A^{(s)} = - {\alpha_3 \over \alpha_r}\, C \delta^{rs}
\quad r,s = 1,2
\end{equation}
\begin{equation}
A^{(r) T} C^{-1} A^{(s)} = - {\alpha_r \over \alpha_3}\, C^{-1}
\delta^{rs} \quad r,s = 1,2 .
\end{equation}
The symbol $T$ means matrix transpose. Some additional useful
identities are
\begin{equation}
\sum_{r=1}^3 {1 \over \alpha_r} A^{(r)} C A^{(r)T} =0,
\end{equation}
\begin{equation}\label{AAT2}
\sum_{r=1}^3 \alpha_r A^{(r)} C^{-1} A^{(r)T} = {\alpha \over 2} B
B^T,
\end{equation}
where we have introduced
\begin{equation}
\alpha = \alpha_1 \alpha_2 \alpha_3.
\end{equation}

Additional matrices that involve the mass parameter $\mu$ of the
plane-wave geometry, introduced in \cite{Spradlin:2002ar}, are
\begin{equation}
(C_r)_{mn} = \omega_{rm} \delta_{mn} = \sqrt{m^2 + \mu^2
\alpha_r^2}\, \delta_{mn}
\end{equation}
and
\begin{equation}
U_r = C^{-1} (C_r -\mu \alpha_r).
\end{equation}
Note that
\begin{equation} \label{Uformula}
(U_r)^{-1} = C^{-1} (C_r + \mu \alpha_r) = U_r + 2 \mu \alpha_r
C^{-1}.
\end{equation}
A crucial construct is the infinite matrix
\begin{equation}
\Gamma_+ = \sum_{r=1}^3 A^{(r)} U_r A^{(r)T}.
\end{equation}
Explicit formulas for the inverse of $\Gamma_+$ are a main goal of
our work. Related quantities that also are needed are the infinite
vector
\begin{equation}
Y = \Gamma_+^{-1} B
\end{equation}
and the scalar
\begin{equation}
k = B^T \Gamma_+^{-1} B.
\end{equation}

\section*{Appendix B. Derivation of the differential equation}

This appendix sketches the derivation of eq.~(\ref{diffeq}), which we copy here
\begin{equation} \label{diffeq2}
\frac{\partial Y_m}{\partial \mu} = \left[\frac{1}{2} \frac{\partial
F}{\partial \mu} \left(1 - \frac{\mu}{\omega_m}\right) -
\frac{\mu}{\omega_m^2}\right] Y_m .
\end{equation}

The matrix $\Gamma_+ = \sum_r A^{(r)} U_r A^{(r){\rm T}}$, introduced in
Appendix~A, only
depends on $\mu$ through the dependence of $U_r$ on $\mu$. Its
derivative can be written in the form
\begin{equation}
\frac{\partial \Gamma_+}{\partial \mu} = - \frac{1}{2} \alpha B
B^{\rm T}  + \mu N,
\end{equation}
where
\begin{equation}
N = \sum_{r=1}^3  \alpha_r^2 A^{(r)} C^{-1} C_r^{-1} A^{(r){\rm T}}.
\end{equation}
It follows that
\begin{equation}
\label{follows}
\frac{\partial Y}{\partial \mu} =  \frac{1}{2} k \alpha Y  - \mu
\Gamma_+^{-1} NY.
\end{equation}
The product $NY$ can be recast in the form
\begin{equation}
\label{recast}
 NY = g_1 C_3^{-2} B + g_2 B,
\end{equation}
where we define the coefficients $g_1$ and $g_2$ to be the scalar quantities
\begin{equation}
\label{gdefs}
g_1 = \frac{2(1 + \mu \alpha k)}{2 + \mu \alpha k + \mu^2 \alpha
k_1},
\end{equation}
\begin{equation}
g_2 = \left(\frac{\alpha}{2}\right) \frac{\alpha k^2  + \mu \alpha k
k_1 + 2 k_1}{2 + \mu \alpha k + \mu^2 \alpha k_1},
\label{gonegtwo}
\end{equation}
and 
\begin{equation}
k_i = B^{\rm T} C_3 ^{-i} Y.
\label{ki}
\end{equation}

The above equations imply that
\begin{equation}\label{keq}
\frac{\partial k}{\partial \mu} = B^{\rm T} \frac{\partial Y}{\partial
\mu} = \frac{1}{2} \alpha k^2 - \mu g_2 k -\mu g_1 k_2.
\end{equation}
This is not very useful as it stands, since there is no other
apparent way to determine $k_2$. ($k_1$ could be determined, but
that will turn out not to be necessary.) Substituting the equation
for $NY$ and an identity for $[C_3^{-2}, \Gamma_+^{-1}]$
deduced from eq.~(\ref{symformula}), one can recast eq.~(\ref{follows}) in the form
\begin{equation}\label{Yeq}
\frac{\partial Y}{\partial \mu} =  (F_0 +  F_1 C_3^{-1} + F_2
C_3^{-2}) Y,
\end{equation}
where the scalar functions $F_i$ are given by
\begin{equation}
F_0 = \frac{1}{2} \alpha k - \mu g_2 + \frac{1}{2} \mu g_1 \frac{\alpha
}{1 + \mu \alpha k} (k_1 - \mu k_2),
\end{equation}
\begin{equation}
F_1 = -\frac{1}{2}  \mu g_1 \frac{\alpha}{1 + \mu \alpha k} (k -
\mu^2 k_2),
\end{equation}
\begin{equation}
F_2 =  - \mu g_1 + \frac{1}{2} \mu^2 g_1 \frac{\alpha}{1 + \mu
\alpha k} (k - \mu k_1).
\end{equation}
Using the equations above to eliminate $g_1$, $g_2$, $k_1$, and $k_2$, we find
\begin{equation}
F_2 = - \mu, \qquad F_1 = - \mu F_0, \qquad F_0 =
\frac{\alpha}{2} \frac{1}{1 + \mu \alpha k} ( k + \mu k').
\end{equation}
This allows us to rewrite eq.~(\ref{Yeq}) in the desired form eq.~(\ref{diffeq2}).

\section*{Appendix C. The integral transform}

The analysis in Sec.~4 of the text required solving the following integral
equation for $f(x)$
\begin{equation} \label{gform}
g(z) = \int_0^\infty
\frac{f(x)}{(x^2 + z^2)^{3/2}} dx,
\end{equation}
where $g(z)$ is a given function that is holomorphic in the
right-half plane and vanishes at infinity in that half plane. We
claim~\cite{He:2002zu} that the unique solution is
\begin{equation} \label{fform}
f(x) = i \frac{x^2}{\pi} \int_0^{\pi} g(i x \cos \theta) \cos
\theta \, d \theta.
\end{equation}

The proof that elimination of $f$ from eqs.~(\ref{gform}) and
(\ref{fform}) gives $g=g$ is a consequence of elementary
integration. This proves the existence of a solution. The proof of
uniqueness requires that elimination of $g$ should give $f=f$.
After making changes of variables and deforming integration
contours, one can argue that this requires the identity
\begin{equation}
\delta(y-y') = \frac{\sqrt{y}}{4 \pi i} \int_{\cal C}
\frac{dw }{\sqrt{1 + w}} \frac{1}{(w y + y')^{3/2}},
\end{equation}
where $y, y' > 0$. The contour can be taken to be the unit circle $|z| = 1$,
in the counterclockwise sense, starting and ending at the point $z=-1$. It is
an elementary application of Cauchy's theorem to show that this integral vanishes
for $y < y'$ and for $y' < y$. That it is has the right singularity at $y=y'$
can be verified by showing that the Laplace transform of both sides give an identity.


\newpage

\begin{thebibliography}{99}

\bibitem{Blau:2001ne}
M.~Blau, J.~Figueroa-O'Farrill, C.~Hull and G.~Papadopoulos, ``A
new maximally supersymmetric background of IIB superstring
theory,'' JHEP {\bf 0201}, 047 (2002) [arXiv:hep-th/0110242].

\bibitem{Metsaev:2001bj}
R.~R.~Metsaev, ``Type IIB Green-Schwarz superstring in plane wave
Ramond-Ramond  background,'' Nucl.\ Phys.\ B {\bf 625}, 70 (2002)
[arXiv:hep-th/0112044].

\bibitem{Berenstein:2002jq}
D.~Berenstein, J.~M.~Maldacena and H.~Nastase,
``Strings in flat space and pp waves from N = 4 super Yang Mills,''
JHEP {\bf 0204}, 013 (2002)
[arXiv:hep-th/0202021].

\bibitem{Spradlin:2002ar}
M.~Spradlin and A.~Volovich,
``Superstring interactions in a pp-wave background,''
Phys.\ Rev.\ D {\bf 66}, 086004 (2002)
[arXiv:hep-th/0204146].

\bibitem{Spradlin:2002rv}
M.~Spradlin and A.~Volovich,
``Superstring interactions in a pp-wave background. II,''
JHEP {\bf 0301}, 036 (2003)
[arXiv:hep-th/0206073].

\bibitem{Pankiewicz:2002tg}
A.~Pankiewicz and B.~Stefa\'nski,
``pp-wave light-cone superstring field theory,''
Nucl.\ Phys.\ B {\bf 657}, 79 (2003)
[arXiv:hep-th/0210246].

\bibitem{Green:1982tc}
M.~B.~Green and J.~H.~Schwarz, ``Superstring Interactions,''
Nucl.\ Phys.\ B {\bf 218}, 43 (1983).

\bibitem{Green:hw}
M.~B.~Green, J.~H.~Schwarz and L.~Brink, ``Superfield Theory of
Type II Superstrings,'' Nucl.\ Phys.\ B {\bf 219}, 437 (1983).

\bibitem{Pankiewicz:2003pg}
A.~Pankiewicz,
``Strings in plane wave backgrounds,''
Fortsch.\ Phys.\  {\bf 51}, 1139 (2003)
[arXiv:hep-th/0307027].

\bibitem{Plefka:2003nb}
J.~C.~Plefka,
``Lectures on the plane-wave string / gauge theory duality,''
arXiv:hep-th/0307101.

\bibitem{Maldacena:2003nj}
J.~M.~Maldacena, ``TASI 2003 lectures on AdS/CFT,''
arXiv:hep-th/0309246.

\bibitem{Spradlin:2003xc}
M.~Spradlin and A.~Volovich,
``Light-cone string field theory in a plane wave,''
arXiv:hep-th/0310033.

\bibitem{Sadri:2003pr}
D.~Sadri and M.~M.~Sheikh-Jabbari,
``The plane-wave / super Yang-Mills duality,''
arXiv:hep-th/0310119.

\bibitem{Schwarz:2002bc}
J.~H.~Schwarz,
``Comments on superstring interactions in a plane-wave background,''
JHEP {\bf 0209}, 058 (2002)
[arXiv:hep-th/0208179].

\bibitem{Pankiewicz:2002gs}
A.~Pankiewicz, ``More comments on superstring interactions in the
pp-wave background,'' JHEP {\bf 0209}, 056 (2002)
[arXiv:hep-th/0208209].

\bibitem{He:2002zu}
Y.~H.~He, J.~H.~Schwarz, M.~Spradlin and A.~Volovich,
``Explicit formulas for Neumann coefficients in the plane-wave geometry,''
Phys.\ Rev.\ D {\bf 67}, 086005 (2003)
[arXiv:hep-th/0211198].

\bibitem{Klebanov:2002mp}
I.~R.~Klebanov, M.~Spradlin and A.~Volovich,
``New effects in gauge theory from pp-wave superstrings,''
Phys.\ Lett.\ B {\bf 548}, 111 (2002)
[arXiv:hep-th/0206221].


\end{thebibliography}
\end{document}